\documentclass[a4paper, 10 pt, conference]{ieeeconf} 
\IEEEoverridecommandlockouts
\usepackage{cite}
\usepackage{amsmath,amssymb,amsfonts}
\usepackage{algorithmic}
\usepackage{graphicx}
\usepackage{textcomp}
\usepackage{xcolor}
\usepackage{bm}
\newtheorem{proposition}{Proposition}

\def\BibTeX{{\rm B\kern-.05em{\sc i\kern-.025em b}\kern-.08em
    T\kern-.1667em\lower.7ex\hbox{E}\kern-.125emX}}
\begin{document}

\title{Feature Construction and Selection for PV Solar Power Modeling \\
}

\author{Yu Yang$^{1}$, Jia Mao$^{1}$, Richard Nguyen$^{2}$, Annas Tohmeh$^{2}$, Hen-Geul Yeh$^{2}$
	\thanks{$^{1}$Yu Yang and Jia Mao are with Department of Chemical Engineering, California State University Long Beach
		{\tt\small yu.yang@csulb.edu}}%
	\thanks{$^{2}$Richard Nguyen, Annas Tohmeh, Hen-Geul Yeh are with the Department of Electrical Engineering, California State University Long Beach}
}

\maketitle

\begin{abstract}
Using solar power in the process industry can reduce greenhouse gas emissions and make the production process more sustainable. However, the intermittent nature of solar power renders its usage challenging. Building a model to predict photovoltaic (PV) power generation allows decision-makers to hedge energy shortages and further design proper operations. The solar power output is time-series data dependent on many factors, such as irradiance and weather. A machine learning framework for 1-hour ahead solar power prediction is developed in this paper based on the historical data. Our method extends the input dataset into higher dimensional Chebyshev polynomial space. Then, a feature selection scheme is developed with constrained linear regression to construct the predictor for different weather types. Several tests show that the proposed approach yields lower mean squared error than classical machine learning methods, such as support vector machine (SVM), random forest (RF), and gradient boosting decision tree (GBDT).
\end{abstract}


\section{Introduction}
Solar power becomes increasingly important in the energy market as more PV panels are installed globally. However, the intermittent character of solar incurs a great safety issue  because the high fluctuation of energy supply leads to the system instability and may damage connected appliances. Hence, an accurate prediction model of solar power generation is highly desired for the system integration and control~\cite{LVS:15, CaB:12}.           

Solar power prediction approaches based on the physical, statistical, and machine learning models have been proposed. The physical models use either previous observations or numerical weather prediction (NWPs) as inputs. For example, the persistence (PSS) model simply assumes that the current power generation is equal to the previous one. The total sky imager (TSI) model relies on the image processing technique and cloud tracking for 15-30 minutes ahead prediction \cite{CUL:11}. These two methods are limited to the short horizon prediction because the cloud cover may change rapidly. Many types of clear sky model can be used to estimate the solar irradiance \cite{AUP:19}, which then is inputted to a solar PV modeling algorithm for power prediction~\cite{Cla:17}. The statistical approach is extensively studied in \cite{BMN:09}, which uses autoregressive (AR) for the short-term prediction and AR with exogenous input (ARX) for the long-term forecast. In \cite{ChR:87}, a clear sky model is introduced to normalize the solar power data, and then an auto-regressive integrated moving average (ARIMA) model is built for the stochastic cloud cover. A probabilistic model is developed in \cite{LSL:20} to determine the joint distribution of hourly-ahead horizontal irradiation and measured solar power supply. Many machine learning methods are applied for solar power prediction and show significant superiority over the traditional physical and statistical approaches. The neural network (NN) model has become very popular since the 1990s~\cite{IPC:13}. The forecast models based on linear, feed-forward, recurrent, and radial basis NN have been developed \cite{SfC:00, HAR:01, CaL:08} for the global horizontal irradiance (GHI). Other machine learning methods, such as support vector machines (SVM) using multiple kernels \cite{SSI:11}, are also available in literature. A comprehensive comparison study on the day-ahead hourly forecast of solar power generation is presented in \cite{GBC:18}, where the second-order grey-box regression method, NN, quantile random forest (RF), k-Nearest Neighbors (kNN), and support vector regression (SVR) are investigated. Their results show that these approaches have similar overall accuracy. An ensemble average is proposed to synthesize these methods and achieve the best performance under all weather conditions.         

The contribution of this paper is building a regression model based on the high-order basis functions for 1-hour ahead solar power prediction. The weather conditions, temperature, dew point, humidity, and wind speed are inputs to the predictor. The one-step (15-minute) past solar generation is introduced as an autoregressive term in the model. An essential innovation of this work is introducing Chebyshev polynomials and trigonometric functions into the regression model to form a higher dimensional feature space. Then, a wrapper method is employed to select suitable features for different weather conditions. Based on the selected features, a constrained least squares problem is solved to determine the model coefficients. In case studies, we show that the proposed approach is more accurate than SVR, RF, and gradient boosting decision tree (GBDT).

The remainder of this paper is organized as follows. The background knowledge and dataset are shown in Section 2. In Section 3, the regression model and feature selections are presented. In Section 4, several classical machine learning methods are implemented through scikit-learn package and compared with our method. The conclusion is drawn in the final Section 5.

\section{Background and Data}
This work aims to predict 1-hour ahead solar power generation using weather data. Only short-horizon prediction is studied because long-range weather forecast may not be accurate, especially when the cloud cover plays a very important role in the output. The Long Beach, California weather record from January to June 2014 is gathered to extract the information of temperature, dew point, humidity, wind speed, and weather type. Different from \cite{SSI:11}, precipitation is not considered as a useful feature since it does not vary sufficiently in California within the entire day. Regarding the weather type, we combine cloudy, mostly cloudy, and partly cloudy as one group. The haze, fog, and blowing dust are in the same group because they are all classified as horizontal obscuration. Therefore, three weather types are considered, including cloudy, fair, and haze. Here we do not study the rainy weather because no such data is available during the daytime at the studied location.     

The California Solar Initiative (CSI) 15-minute interval data is used as the solar energy output. This dataset was built on the measured production data from 414 of the 504 solar systems and further improved by simulation for missing data and reflection of the true character of the system \cite{CSI}. An important step is to match the location and timestamp of the solar power production and weather data. Here a PV system in Long Beach airport is studied, where the historical weather data is available with details. Because the sampling time of weather data is irregular, we align each solar power data item with its closest weather record. The solar power prediction models can be expressed as:
\begin{align}
	y(k+1)=F_{[i]}(y(k), \bm{u}(k))
\end{align}         
where $k$ represents the sampling time instant with 15 minutes interval; $y$ denotes the solar power output; $\bm{u}=[\textrm{temperature, dew point, humidity, wind speed}]$; $F_{[i]}$ is the predictor function. Note that the weather type is not directly used as the input because associated cloud coverage is not quantified in the record. Instead, we use subscript $[i]$ to denote different weather type and develop their models separately. In prediction, the employed model can be switched based on the weather type at time instant $k$. Here $y(k)$ in the regression function is an autoregressive term to take the most recent measurement into account.

Considering that the sun elevation and azimuth vary month by month, the training, validation, and testing dataset may not cross a long period. Six datasets are designed and shown in Table~\ref{Tab:Data_Date}. The training set has 25 days, whereas validation and testing datasets all have 5 days.

\begin{table}[htbp]
	\centering
	\caption{Date of training, validation and testing dataset} \label{Tab:Data_Date}
	\begin{tabular}{|cccc|}
		\hline
	               	& Training  &  	Validation & Testing\\
		\hline
	    Dataset 1  & Jan 1-25  & Jan 26-30 & Jan 31-Feb 4\\ 
	    Dataset 2  & Feb 1-25  & Feb 26-Mar 2 & Mar 3-7 \\
	    Dataset 3  & Mar 1-25 & Mar 26-30 &  Mar 31-Apr 4 \\
	    Dataset 4  & Apr 1-25 &  April 26-30  &  May 1-5   \\
	    Dataset 5  &   Apr 26-May 20 & May 21- 25  &  May 26-30 \\
	    Dataset 6  & May 20-Jun 13  &  Jun 14-18  & Jun 19-23\\
		\hline
	\end{tabular}
\end{table}   

\begin{figure}[!h]
\centering
\includegraphics[width=3.4in, height=2.65in]{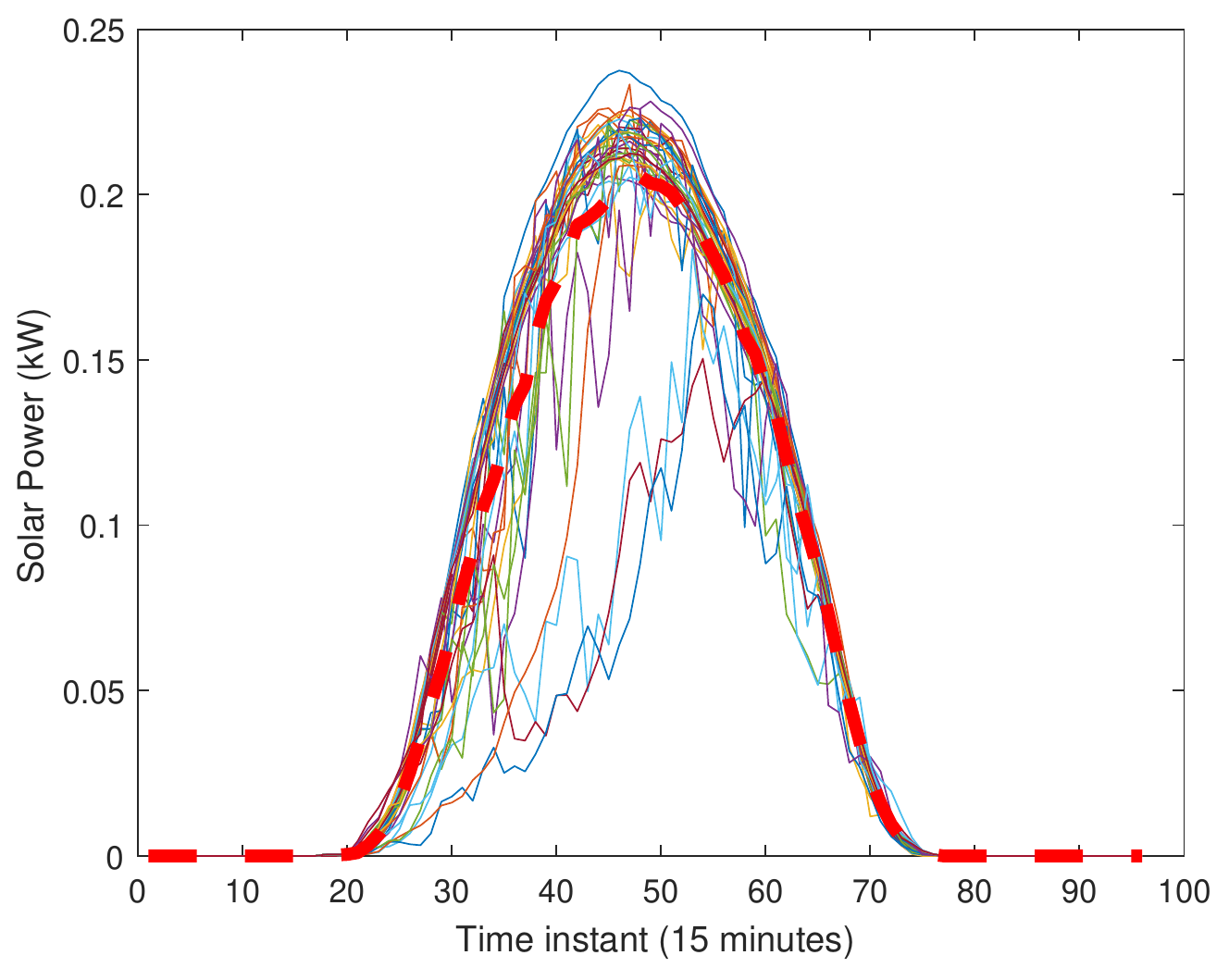}
\caption{Each curve represents a daily solar power profile.}
\label{fig:May30_solar}
\end{figure}

A simple model can be built upon the mean of power generation profiles in the training set, denoted as a positive variable $\bar{y}(k)$. However, such an average may not reflect the actual power dynamics under different weather conditions. In Fig.~\ref{fig:May30_solar}, the solar power profiles of every day in dataset 3 are plotted as an example. The thick dash line represents the mean power output. Large deviations occur when the weather conditions are significantly different from the average.
The mean value obtained from the training dataset is removed to reduce the time dependence. Then, models are developed to predict the deviation $y'=y-\bar{y}$ rather than the raw value $y$, such that the mean squared error (MSE) can be reduced significantly.      

\section{Prediction Model}
Starting from the easiest AR model, we denote the following basic regressor: 
\begin{align}
\phi(k+1)=[y'(k), \bm{u}_{1}(k), \bm{u}_{2}(k), \bm{u}_{3}(k), \bm{u}_{4}(k), \bm{u}_5(k)]
\end{align}
where $\bm{u}_{1}$ is the temperature; $\bm{u}_{2}$ is the dew point; $\bm{u}_{3}$ is the humidity; $\bm{u}_{4}$ is the wind speed; and $\bm{u}_{5}$ is the time. Here we omit the index of data items. Then, one may assume that output $y'(k+1)$ is linearly dependent on the input $\bm{u}(k)$ and autoregressive term $y'(k)$. However, such a simple model cannot achieve satisfactory performance. Instead, a high-order polynomial model is designed based on $\phi$. Clearly, there are so many options for polynomial terms. One cannot arbitrarily specify the degree and formulas of the polynomial used as features. Therefore, a feature construction and selection procedure is designed to determine which polynomial terms should be chosen.


\subsection{Feature Construction}
Let us briefly describe candidate features. For the deviation outputs $y'(k)$, the bell shape data still can be found in some trials. Thus, the regressor $\phi(k)$ is extended by attaching two more variables: $\bm{u}_6(k)=\cos(\pi \bm{u}_5(k)/24)$ and $\bm{u}_7(k)=\sin(\pi \bm{u}_5(k)/24)$. This extended regressor becomes 
\begin{align}
\phi^*(k+1)=&[y'(k), \bm{u}_{1}(k), \bm{u}_{2}(k), \bm{u}_{3}(k), \bm{u}_{4}(k), \bm{u}_5(k), \nonumber \\ 
         & \bm{u}_6(k), \bm{u}_7(k) ]  \nonumber
\end{align}
Then, the normalized regressor is
\begin{align*}
\tilde{\phi}(k)= \frac{\phi^*(k)}{\bar{\phi}}
\end{align*}
where denominator vector $\bar{\phi}$ can be chosen as the maximum absolute value of each variable in the data such that $\tilde{\phi}(k)$ is within the range $[-1,1]$. The first kind Chebyshev polynomial can be constructed to form the following feature set:
\begin{align*}
\bm{C}_{0}&=1,~\bm{C}_{1}=\tilde{\phi},~\bm{C}_{2}=2\tilde{\phi}^2-1,~\bm{C}_{3}=4\tilde{\phi}^3-3\tilde{\phi},\\
\bm{C}_{4}&=8\tilde{\phi}^4-8\tilde{\phi}^2+1, ~\bm{C}_{5}=16\tilde{\phi}^5-20\tilde{\phi}^3+5\tilde{\phi},\\
\bm{C}_{6}&=32\tilde{\phi}^6-48\tilde{\phi}^4+18\tilde{\phi}^2-1,\\
\bm{C}_{7}&=64\tilde{\phi}^7-112\tilde{\phi}^5+56\tilde{\phi}-7\tilde{\phi},\\
\bm{C}_{8}&=128\tilde{\phi}^8-256\tilde{\phi}^6+160\tilde{\phi}^4-32\tilde{\phi}^2+1,\\
\bm{C}_{9}&=256\tilde{\phi}^9-576\tilde{\phi}^7+432\tilde{\phi}^5-120\tilde{\phi}^3+9\tilde{\phi},\\
\bm{C}_{10}&=512\tilde{\phi}^{10}-1280\tilde{\phi}^8+1120\tilde{\phi}^6-400\tilde{\phi}^4+50\tilde{\phi}^2-1
\end{align*}
Only $\bm{C}_0-\bm{C}_{10}$ are considered because more polynomials may render the feature selection step more computationally expensive. Except $\bm{C}_0$,  all $\bm{C}_w~ \forall w\in\{1,2,\ldots,10\}$ are matrices with 8 columns. The proposed scheme increases the dimension of basis function and thus is able to represent complex dynamics of the system. Moreover, because the cloud cover impacts solar irradiance but is not reflected in existing features, another three inputs $\bm{u}_{8}, \bm{u}_{9}, \bm{u}_{10}$ are further introduced: 
\begin{equation}
\bm{u}_8(k) = \begin{cases}
1 &  \textrm{if cloudy at time instant $k$}\\
0 & \textrm{else}  
\end{cases} 
\end{equation}
\begin{equation}
\bm{u}_{9}(k) = \begin{cases}
1 &  \textrm{if mostly cloudy at time instant $k$}\\
0 & \textrm{else}  
\end{cases} 
\end{equation}
\begin{equation}
\bm{u}_{10}(k) = \begin{cases}
1 &  \textrm{if partly cloudy at time instant $k$}\\
0 & \textrm{else}  
\end{cases} 
\end{equation}
Because the sky cover data is not quantified, one-hot encoding $\bm{u}_{9}, \bm{u}_{10}, \bm{u}_{11}$ can highlight the different types of clouds. The new features are the product of time and cloudy type:
\begin{align*}
\bm{C}_{11}&=\bm{u}_8\bm{u}_5/24\\
\bm{C}_{12}&=\bm{u}_{9}\bm{u}_5/24\\
\bm{C}_{13}&=\bm{u}_{10}\bm{u}_5/24
\end{align*}    
The rational of introducing $\bm{C}_{11}-\bm{C}_{13}$ with time instant is that cloud incurs high deviation on the irradiance at noon, whereas the variation near sunset or sunrise is relatively small. $\bm{C}_{11}-\bm{C}_{13}$ are generated through feature interaction and not linearly dependent with any existing features.


\subsection{Feature Selection}
In the previous section, several features are introduced to form a pool. Including all of them in the regression model can significantly reduce the training error. However, this is not true for the validation and testing datasets. Namely, overfitting may happen if unnecessary features contribute to the training process. To overcome this issue, we employ a sequential forward selection and backward elimination procedure. A subset of features will be chosen for each dataset such that the prediction model performs well on both training and validation set. In fact, a number of feature selection schemes based on the embedded and filter approaches have been proposed for process system analysis \cite{OKG:18, OKP:19, RCS:19, SWH:20}. The wrapper method is employed in this paper because it evaluates the feature set directly based on the data fitting performance. Before discussing the details of feature selection, let us develop the training scheme.    

For the weather type $i$, we build the predictor model $F_{[i]}$ with unknown coefficients $\bm{a}_{[i]}$:
\begin{align}
F_{[i]}=\sum_{\bm{C}_{w,j}\in\Psi_{[i]}}\bm{a}_{[i],w,j} \bm{C}_{w,j}
\end{align} 
where $\Psi_{[i]}$ represents the set of chosen features for weather type $i$; $\bm{C}_{w,j}$ represents $j^{th}$ column in the feature set $\bm{C}_{w}$. A constrained least squares is presented in (\ref{eq:LS_obj}) to obtain $\bm{a}_{[i]}$:
\begin{align}\tag{$\mathcal{LS}$}
\min_{\bm{a}_{[i]}}~& \sum_{k\in\Gamma_{[i]}}(y'(k+1)-\sum_{\bm{C}_{w,j}\in\Psi_{[i]}}\bm{a}_{[i],w,j} \bm{C}_{w,j})^2  \label{eq:LS_obj}\\
    \mbox{s.t.} & \sum_{\bm{C}_{w,j}\in\Psi_{[i]}}\bm{a}_{[i],w,j} \bm{C}_{w,j}+\bar{y}(k+1)\geqslant 0 \label{eq:LS_con}
\end{align}
where $\Gamma_{[i]}$ includes all data indexes for weather type $i$ within the training set. The objective function in (\ref{eq:LS_obj}) is to minimize the one-step prediction error. Eq.~(\ref{eq:LS_con}) requires the predicted solar power $\hat{y}(k+1)=\hat{y}'(k+1)+\bar{y}(k+1)$ greater or equal to zero.

\begin{figure}[!h]
	\centering
	\includegraphics[width=3.4in, height=2.5in]{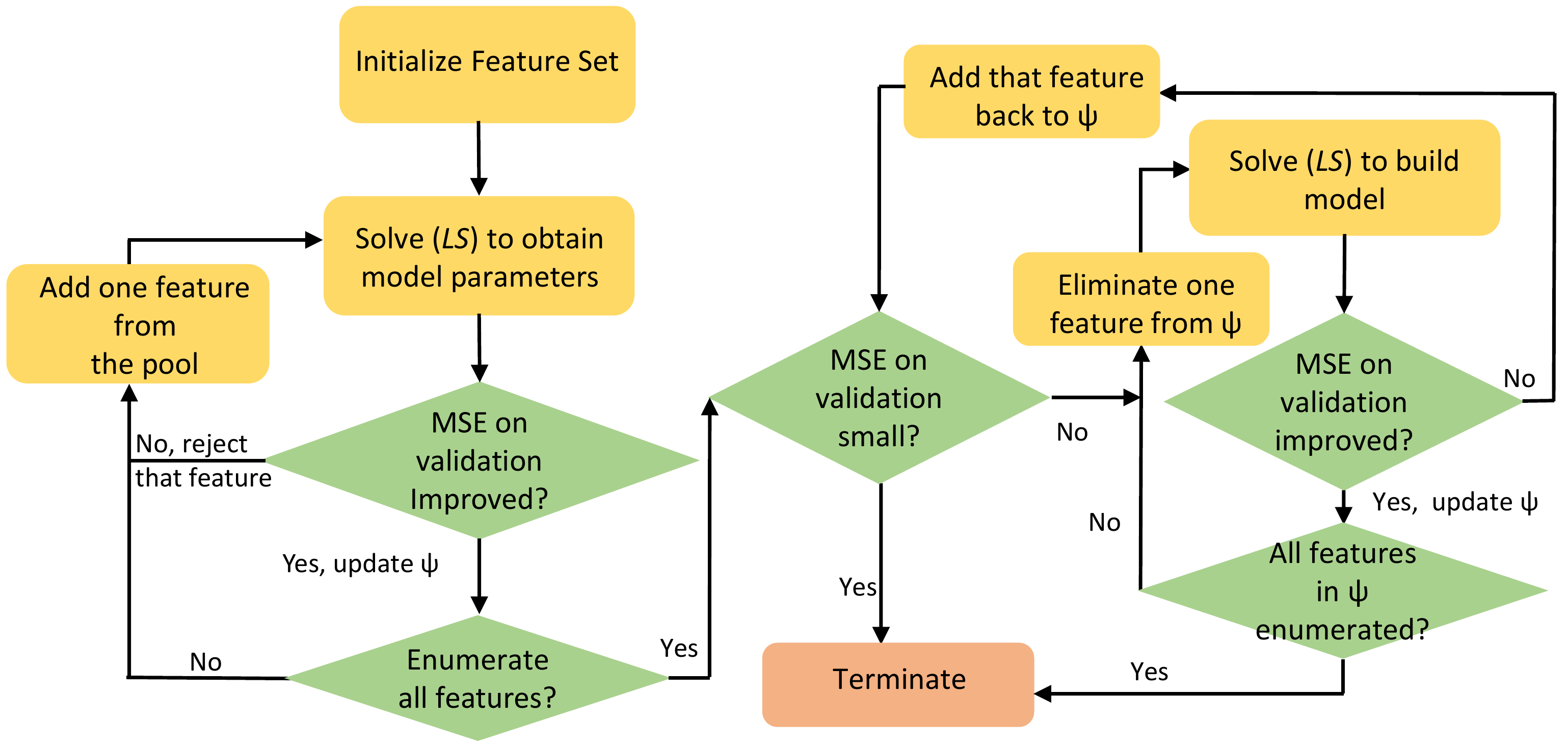}
	\caption{Algorithm 1: Feature selection and elimination for weather $i$.}
	\label{fig:FSBE}
\end{figure}

Then, the flowchart of feature selection and elimination (Algorithm 1) for weather $i$ is shown in Fig.~\ref{fig:FSBE}. The mean squared error (MSE) is the performance index, and the outcome is feature set $\Psi_{[i]}$ with model parameters $\bm{a}_{[i]}$. Here we identify $\bm{a}_{[i]}$ based on the training dataset and the feature evaluation is based on the validation dataset. 

Several comments about Algorithm 1 are in order. \textbf{First}, (\ref{eq:LS_obj}) is a convex optimization problem with at most $10\times 8+4=84$ decision variables, and thereby can be solved quickly. \textbf{Second}, Algorithm 1 is a wrapper method directly using MSE on validation set as a criterion to incorporate or eliminate features. Our method only finds a sub-optimal solution, whereas the global optimality cannot be guaranteed. A possible improvement can be achieved by using Bayesian optimization to choose features. \textbf{Third}, Algorithm 1 only solves (\ref{eq:LS_obj}) to minimize one-step ahead prediction error. Directly minimizing the multi-step prediction error will lead to a high-order nonlinear optimization problem and all different weather types should be considered simultaneously. The resulting computational burden makes feature selection scheme extremely inefficient. \textbf{Fourth}, the selected feature set can be distinct for different datasets and weather types because the sun irradiation and cloud coverage may vary month by month.          

\begin{figure}[!h]
	\centering
	\includegraphics[width=3in, height=2.2in]{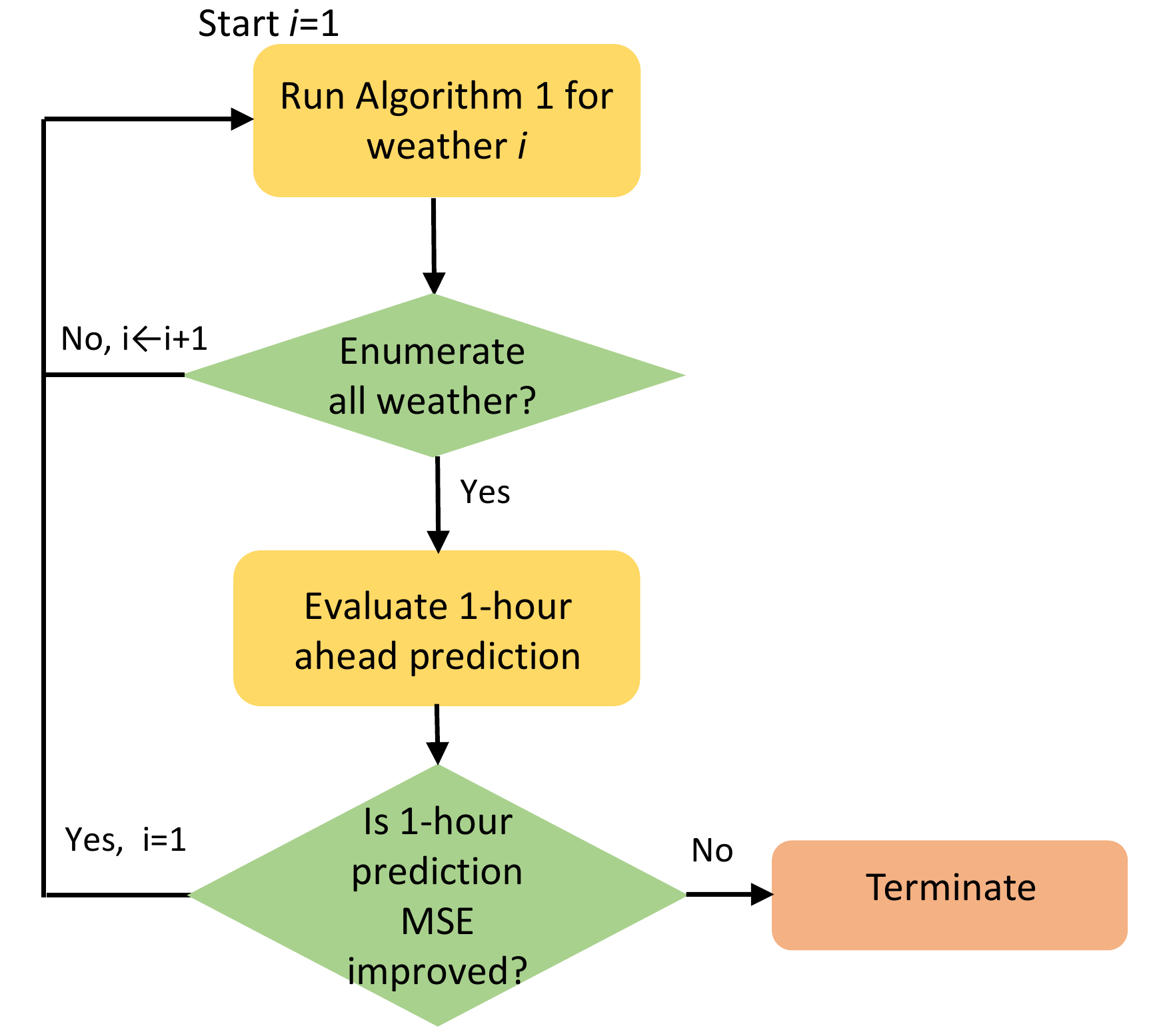}
	\caption{Algorithm 2: Choose the model based on multistep prediction error.}
	\label{fig:A2}
\end{figure}

Next, Algorithm 2 in Fig.~\ref{fig:A2} combines all weather types model together and evaluate the multi-step prediction error on the validation set. This algorithm continuously updates the feature subset for all types of weather until no improvement can be achieved. 

The major difference between multi-step and one-step prediction is that we need to replace the autoregressive term by the prediction value. At time $k$, for one-step prediction $\hat{y}'(k+1|k)$, we use $y'(k|k)$, which is the measurement value. However, for $M$-step prediction $\hat{y}'(k+M|k)$, we need to use $\hat{y}'(k+M-1|k)$, which is based on the previous prediction. It implies that the prediction error at any step will be accumulated and impact the forecast in future steps. 

A constraint can be embedded into (\ref{eq:LS_obj}) to ensure that the multi-step prediction is bounded. Let us define the normalized M-step regressor as
\begin{align*}
\tilde{\phi}(k+M+1|k)= \frac{\phi^*(k+M+1|k)}{\bar{\phi}}
\end{align*}
where $\phi^*(k+M+1|k)=[\hat{y}'(k+M|k), \bm{u}_{1}(k+M), \ldots, \bm{u}_{7}(k+M)],~\forall M\geqslant 0$.
\begin{proposition}
	If constraint (\ref{eq:LS_con2}) is integrated into (\ref{eq:LS_obj}) and $\bar{\phi}\geqslant 1$, the M-step ahead prediction is bounded.
	\begin{align}
	\sum_{w,j} |a_{[i],w,j}|\leqslant 1 \label{eq:LS_con2}
	\end{align}
\end{proposition}
\textbf{Proof:}\\
	The first-kind Chebyshev polynomial $\bm{C}_0$ to $\bm{C}_{10}$ are within $[-1,1]$ because $\tilde{\phi}(k|k)\in[-1,1]$. Features $\bm{C}_{11}$, $\bm{C}_{12}$ and $\bm{C}_{13}$ are also within $[-1,1]$ based on the definition of $\bm{u}_5$, and $\bm{u}_8$ to $\bm{u}_{10}$. If (\ref{eq:LS_con2}) is embedded into (\ref{eq:LS_obj}), then the absolute value of the one-step ahead prediction is bounded by:
	\begin{align*}
	|\hat{y}'(k+1|k)|=|\sum_{w,j} a_{[i],w,j} \bm{C}_{w,j}|\leqslant \sum_{w,j} |a_{[i],w,j}| |\bm{C}_{w,j}| \leqslant 1 
	\end{align*}
	Given $\bar{\phi}\geqslant 1$ and $|\phi^*_1(k+1|k)|=|\hat{y}'(k+1|k)| \leqslant 1$, there is $\tilde{\phi}(k+1|k)\in[-1,1]$. By repeating this process, we have $|\hat{y}'(k+M|k)|<1$, $\forall M$-step prediction.   	

Proposition 1 shows that when $\bar{\phi}\geqslant 1$, the resulting absolute value of output prediction $\hat{y}'(k+M|k)$ is bounded by 1. Hence, the deviation output $y'(k)$ should be pre-scaled to the range $[-1,1]$ in advance. Then, the prediction error $|\hat{y}'(k+M|k)-y'(k+M)|$ is also bounded. Moreover, (\ref{eq:LS_obj}) is feasible even with (\ref{eq:LS_con2}) embedded because zero vector is always a feasible solution.

\section{Results and Discussion}
Six datasets in Table~\ref{Tab:Data_Date} are modeled using the proposed regression method with feature selection. For comparison, we also build SVR, RF, and GBDT models to predict $y'(k)$ for each weather type based on the basic regressor $\phi(k)$. These classical approaches are implemented using scikit-learn package 0.24. The GBDT is implemented through LightGBM~\cite{KMF:17}.   

Here the model performance is evaluated based on the MSE of 1-4 steps prediction, shown in (\ref{eq:MSE}),
\begin{align}
\textrm{MSE}=\frac{\sum_{k=1}^{N-3}\sum_{h=1}^4 (\hat{y}(k+h|k)-y(k+h))^2}{4(N-3)} \label{eq:MSE}
\end{align}
where $N$ is the number of data instances in a dataset. All model parameters are identified through minimizing the predication error on the training set. The hyperparameters of compared approaches (SVR, RF, LightGBM) are tuned by assessing the model performance on validation datasets. Here  data shuffling and cross validation are not implemented because following chronological order is important to the application of this predictor. For SVR, its hyperparameters, including regularization, number of support vectors, type of kernel functions, are tuned through a library function GridSearchCV in scikit-learn. For RF, its hyperparameters are number of trees, maximum number of features for splitting, maximum number of tree levels, minimal number of data points before node splitting, and minimal number of data points in a leaf. For LightGBM, its hyperparameters are tuned using the Fast and Lightweight AutoML (FLAML)~\cite{WWW:21}. In addition, the max depth of lightGBM is also tuned to achieve better performance. Finally, the testing dataset is used to compare the true performance of all considered methods. 

Algorithms 1 and 2 are implemented to select features for our model. Some of high-order polynomial features in $\bm{C}_0-\bm{C}_{10}$ are selected for fair and haze weather. For cloudy weather, besides $\bm{C}_0-\bm{C}_{10}$,  $\bm{C}_{11}-\bm{C}_{13}$ are also chosen by algorithms to construct prediction models.  
  
\begin{figure}[!h]
	\centering
	\includegraphics[width=3.4in, height=2.65in]{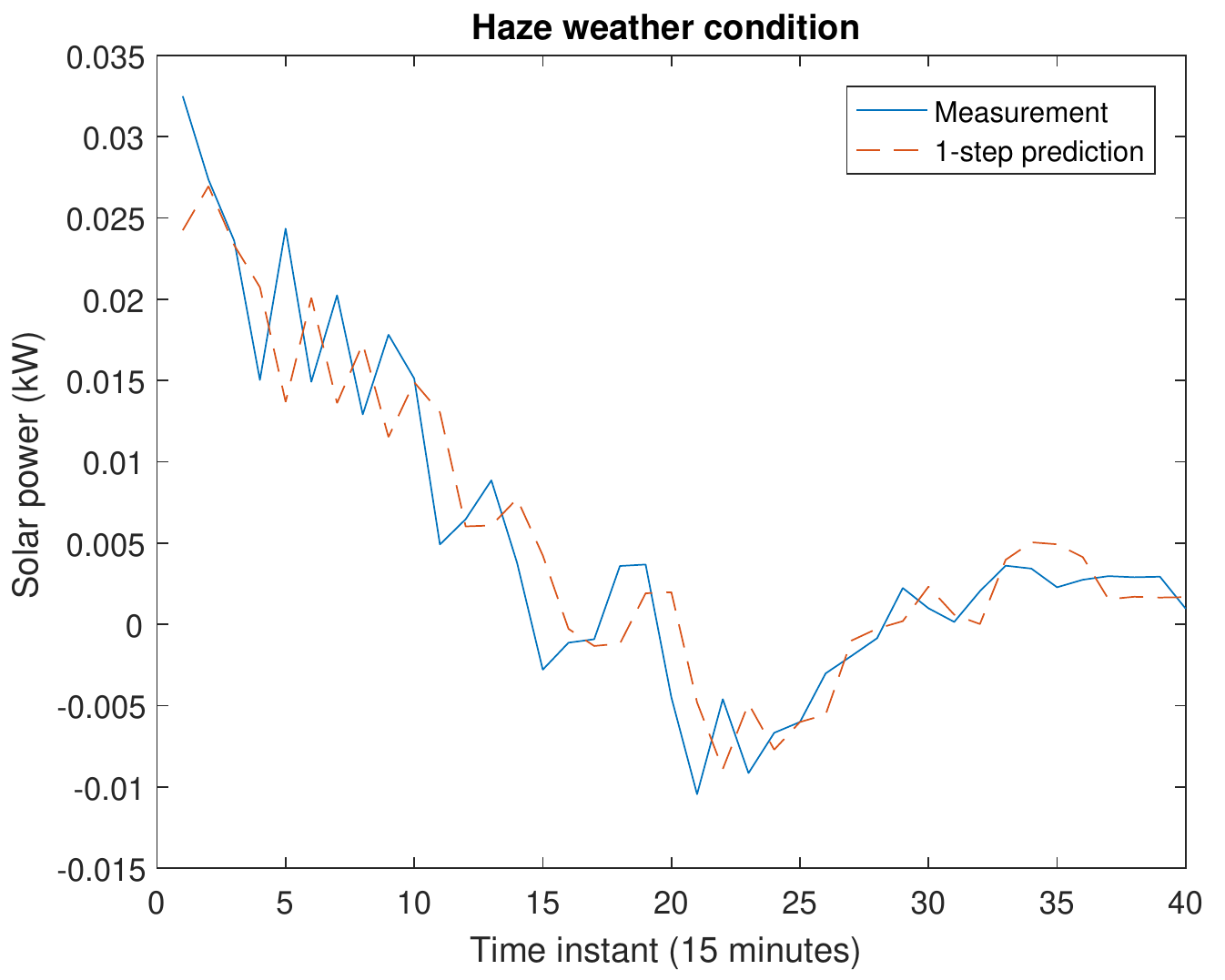}
	\caption{1-step ahead predictions on training set 4 haze weather.}
	\label{fig:5_5_prediction_haze}
\end{figure}

\begin{figure}[!h]
	\centering
	\includegraphics[width=3.4in, height=2.65in]{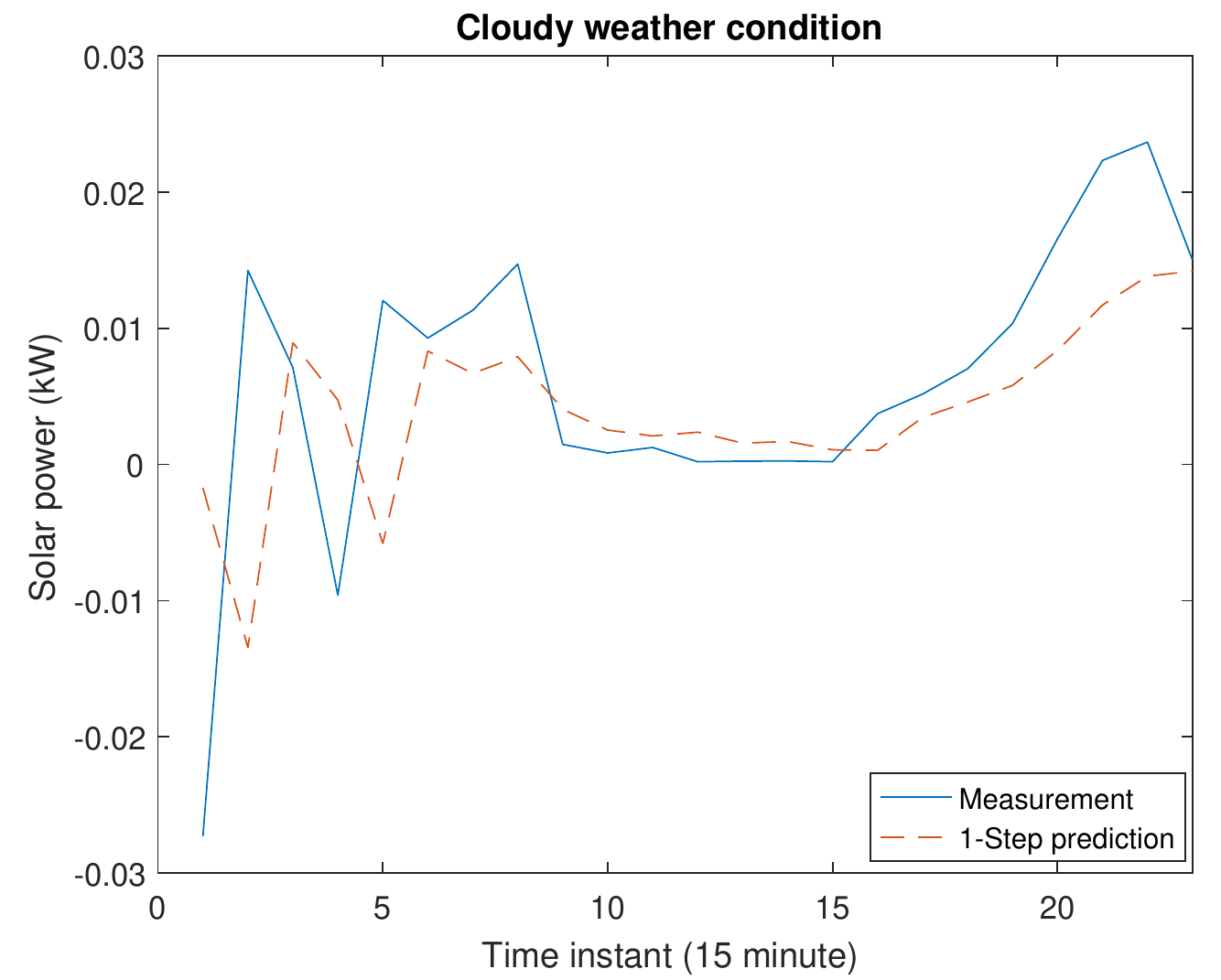}
	\caption{1-step ahead predictions on training set 4 cloudy weather.}
	\label{fig:5_5_prediction_cloudy}
\end{figure}

\begin{figure}[!h]
	\centering
	\includegraphics[width=3.4in, height=2.65in]{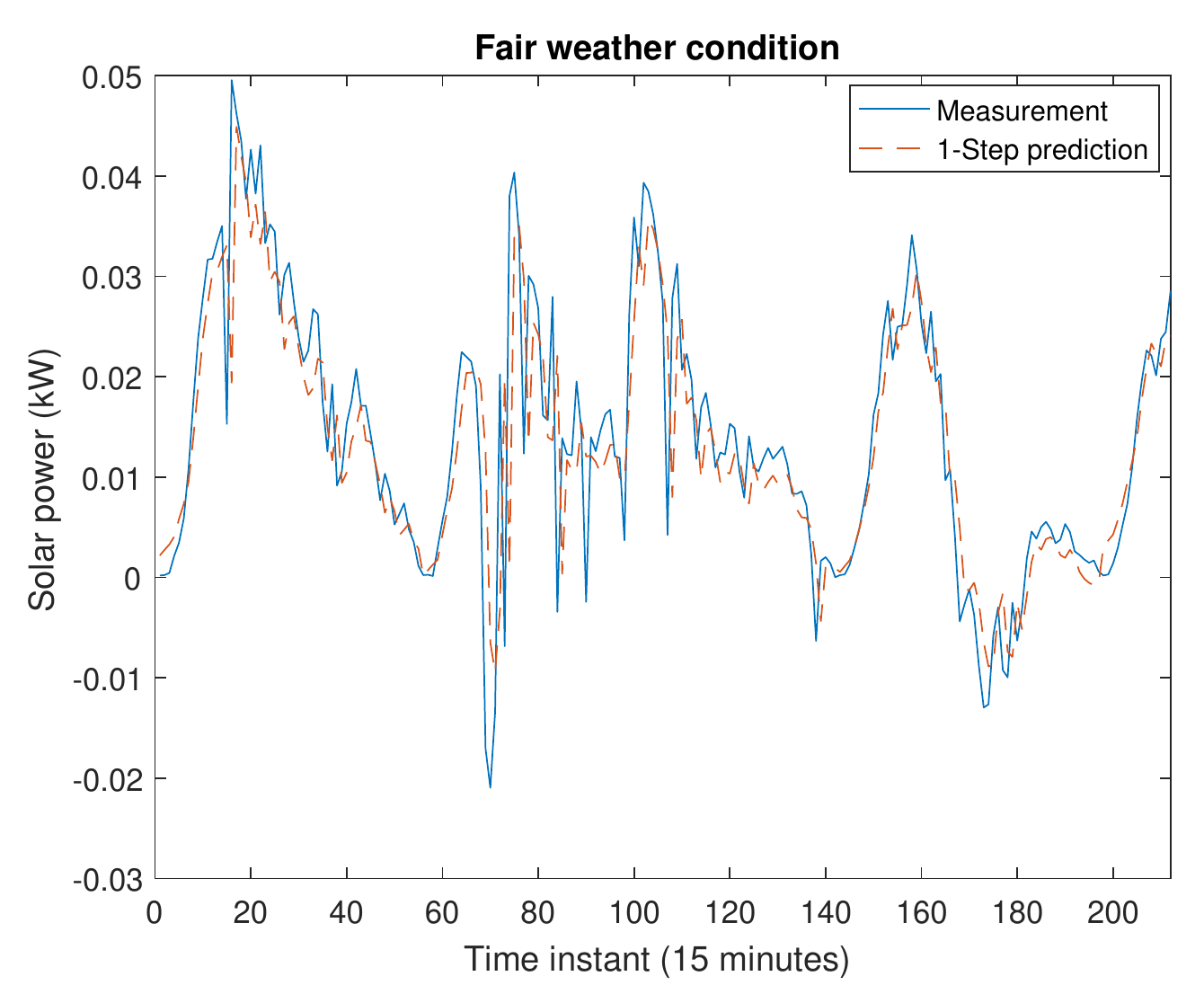}
	\caption{1-step ahead predictions on training set 4 fair weather.}
	\label{fig:5_5_prediction_fair}
\end{figure}

Dataset 4 is used as an example to illustrate the training results. The one-step prediction on Figs.~\ref{fig:5_5_prediction_haze}-\ref{fig:5_5_prediction_fair} shows that the proposed regression model achieves high accuracy for fair and haze weather but is less accurate for cloudy days. The same observation can be found for other datasets. The cloud coverage highly impacts the solar power generation and a quantitative description of cloud could be more helpful in the future research. 

Tables~\ref{Tab:Validation}-\ref{Tab:Test} present combined 1-4 step prediction errors, shown in (\ref{eq:MSE}), for each dataset. Table~\ref{Tab:Validation} shows the MSE on each validation dataset. Our method minimizes MSE on the validation set via feature selection, whereas other methods do this by tuning different hyperparameters. GridSearchCV enables SVR to achieve the best performance on validation datasets through exhaustive search. However, overly tuning hyperparameters may degrade the predictive generality.    

The MSE on each testing dataset is a more important performance index to all data-driven models. Table~\ref{Tab:Test} shows that the proposed model is only slightly worse than SVR on datasets 2,  but is much better than all considered classical machine learning methods on datasets 1, 3, 4, 5 and 6. This result implies that carefully designing and selection feature set for a simple regression model may lead to better results than advanced machine learning methods based on raw features. The constrained least squares (\ref{eq:LS_obj}) can be solved rapidly, and thus is suitable for the proposed feature selection procedure. A future investigation may combine feature selection with SVR to improve its prediction performance.  

\begin{figure}[!h]
	\centering
	\includegraphics[width=3.4in, height=2.65in]{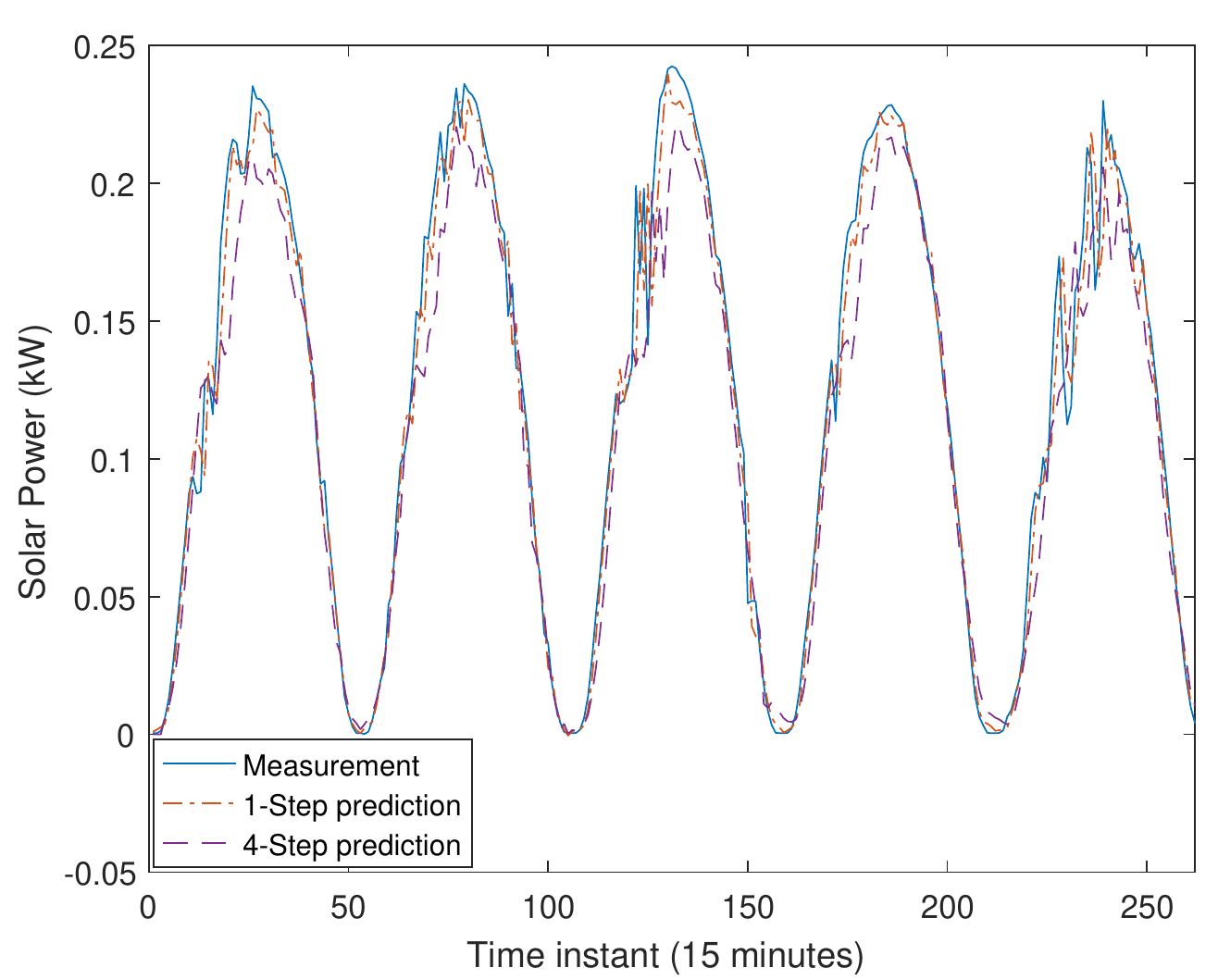}
	\caption{Solar power 1-step and 4-step ahead predictions on testing set 3.}
	\label{fig:4_5_prediction}
\end{figure}
\begin{figure}[!h]
	\centering
	\includegraphics[width=3.4in, height=2.65in]{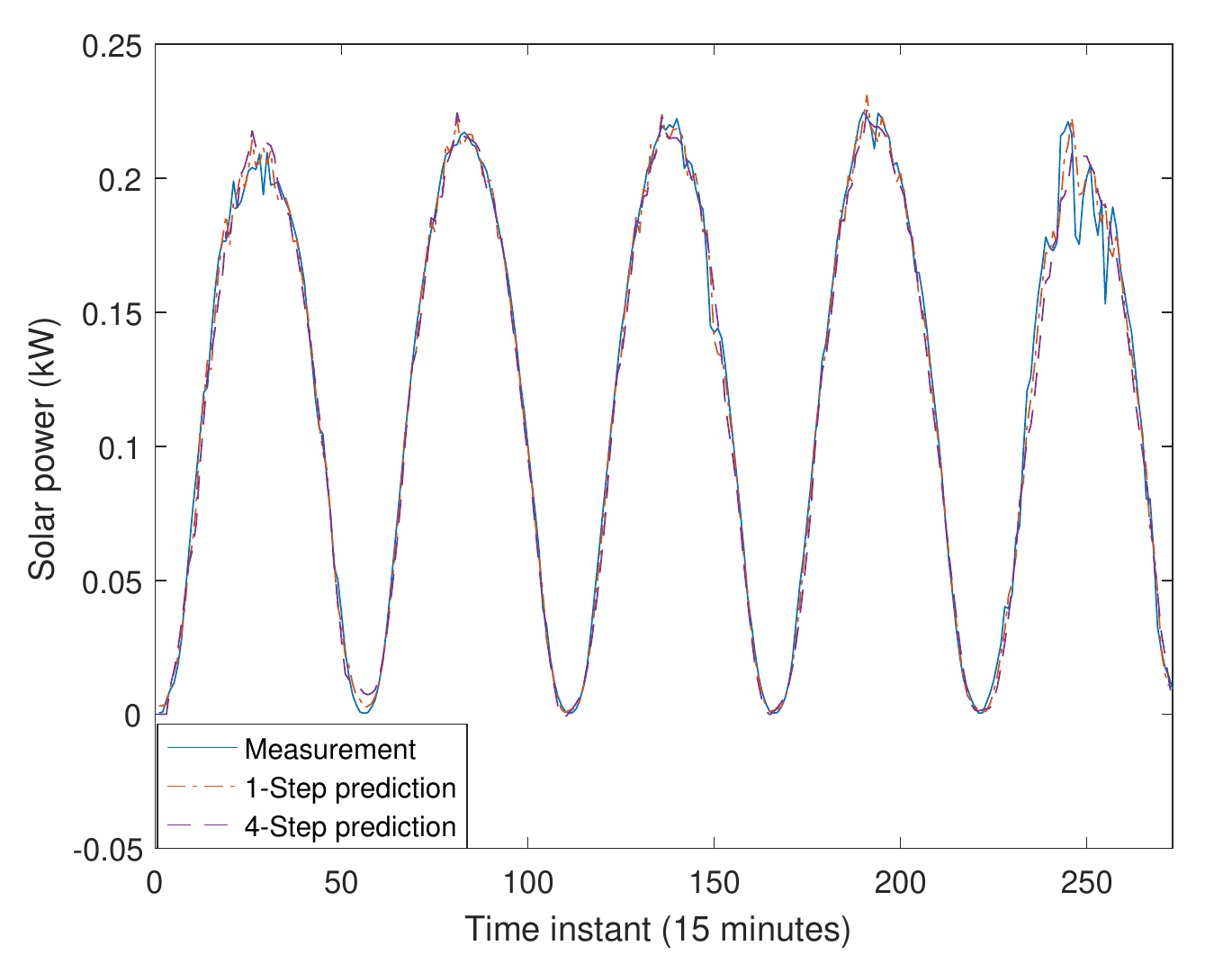}
	\caption{Solar power 1-step and 4-step ahead predictions on testing set 4.}
	\label{fig:5_5_prediction}
\end{figure}

Figs.~\ref{fig:4_5_prediction}-\ref{fig:5_5_prediction} show the measured and predicted values using the proposed approach on testing datasets 3 and 4. Dataset 3 is chosen because the measured power deviates from $\bar{y}$ with large fluctuations. Dataset 4 includes more fair weather data, and thus is less challenging than dataset 3. The prediction error in Fig.~\ref{fig:4_5_prediction} is obvious, whereas in Fig.~\ref{fig:5_5_prediction} is much smaller. It is not surprising that four-step prediction is less accurate than one-step prediction, but the difference is not too significant. Future work can be done to incorporate multi-step prediction into the model development on the training set.  

\begin{table}[htbp]
	\centering
	\caption{Combine 1-4 steps MSE on the validation dataset} \label{Tab:Validation}
	\begin{tabular}{|ccccc|}
		\hline
	 & Proposed Model  & SVR &  RF & LightGBM \\
		\hline
		Dataset 1 &  2.448e-4 & 2.302e-4 & 15.341e-4 & 13.600e-4\\
		Dataset 2 & 9.335e-4  & 3.810e-4 & 39.769e-4 & 35.705e-4\\
		Dataset 3 & 3.826e-4  & 4.920e-4 & 13.179e-4 & 9.087e-4\\
		Dataset 4 & 0.638e-4  & 0.627e-4    & 2.253e-4  & 1.543e-4 \\
		Dataset 5 & 10.106e-4   & 5.242e-4   &  27.986e-4  &    32.700e-4 \\
		Dataset 6 &  0.844e-4   & 0.825e-4     & 1.298e-4  & 1.179e-4 \\
		\hline
	\end{tabular}
\end{table}   
 

\begin{table}[htbp]
	\centering
	\caption{Combined 1-4 steps MSE on the testing dataset} \label{Tab:Test}
	\begin{tabular}{|ccccc|}
		\hline
		 & Proposed Model  & SVR &  RF & LightGBM \\
		\hline
	    Dataset 1 & 6.532e-4 &   7.026e-4   &  11.898e-4    &  18.233e-4   \\ 
	    Dataset 2 & 5.441e-4  &  5.423e-4   &  10.597e-4   &  7.700e-4    \\
	  	Dataset 3 & 3.007e-4  &  5.722e-4   & 12.662e-4   & 14.026e-4   \\
		Dataset 4 & 0.614e-4  &  2.525e-4   & 2.683e-4   &   2.679e-4   \\
		Dataset 5 & 0.869e-4  &  6.452e-4   &  1.725e-4  &   1.302e-4    \\
		Dataset 6 & 0.546e-4  &  0.987e-4   &  1.601e-4     &    1.314e-4     \\		
		\hline
	\end{tabular}
\end{table}

\section{Conclusion}
A regression model is developed for one-hour ahead solar power prediction based on the weather data. The raw solar power generation data is detrended and combined with temperature, dew point, humidity, wind speed to form a basic feature vector. Next, this basic feature is augmented through Chebyshev polynomial and trigonometric functions. A linear combination model of resulting high-dimensional features is developed, whose coefficients are identified based on the training dataset. The feature space is further refined on the validation dataset, and the boundedness of multi-step prediction is shown. Finally, the proposed method is compared with classical machine learning methods, such as SVR, RF, and GBDT, on several testing datasets to demonstrate its superiority in prediction accuracy.

\end{document}